\documentclass[conference]{IEEEtran}
\IEEEoverridecommandlockouts
\usepackage{cite}
\usepackage{amsmath,amssymb,amsfonts}
\usepackage{algorithmic}
\usepackage{graphicx}
\usepackage{textcomp}
\usepackage{listings}
\usepackage{xcolor}
\usepackage{float}
\usepackage{url}
\usepackage{placeins}
\def\BibTeX{{\rm B\kern-.05em{\sc i\kern-.025em b}\kern-.08em
    T\kern-.1667em\lower.7ex\hbox{E}\kern-.125emX}}

\begin{document}
\title{CBL: Compact Encoding of JSON-LD Data using CBOR and Bitmaps for Web of Things}

\author{
\IEEEauthorblockN{Prudhvi Gudla$^{1}$, Priyanka Rawat$^{2}$  and Kamal Singh$^{3}$}
\IEEEauthorblockA{
\textit{$^{1}$Indian Institute of Technology, Kharagpur, India}\\
\textit{$^{2}$Univ Jean Monnet, IOGS, CNRS, UMR 5516, LaHC, F - 42023 Saint-Etienne, France} \\
\textit{$^{3}$Univ Jean Monnet, Telecom Saint Etienne, F - 42023 Saint-Etienne, France} \\
Email: $^{1}$gvssprudhvi@kgpian.iitkgp.ac.in, $^{2}$priyanka.rawat@univ-st-etienne.fr, $^{3}$kamal.singh@univ-st-etienne.fr}
\thanks{The work by Prudhvi Gudla was accomplished during his internship stay in Laboratory Hubert Curien, Saint-Etienne, France.}
}

\maketitle

\begin{abstract}
%JSON-LD (JavaScript Object Notation for Linked Data) is popular data representation formats used in web development and other applications. Web of Things is a new paradigm which combines web technologies and Internet of Things. However, JSON-LD can be too verbose for constrianed objects. This paper explores the conversion of JSON-LD to dictionary based encoding of terms combined with CBOR encoding. This results in a compact representation which can provide bandwidth as well as energy savings. Results show that such encoding can provide around 80 percent of bandwidth savings. 

%Web of Things (WoT) brings web technologies and knowledge graphs to Internet of Things. JSON-LD, with its popularity in representing and exchanging structured data on the web, can be a suitable data format for WoT. However, its verbose nature can pose challenges for constrained IoT devices with limited bandwidth, and memory. 
The concept of Web of Things (WoT) merges web technologies with knowledge graphs in the context of Internet of Things. Given its widespread adoption in representing and exchanging 
structured data online, JSON-LD could be an effective format for WoT. Nevertheless, its verbose nature may present challenges for resource-constrained IoT devices with limited 
bandwidth and memory capacities.

In this paper, we present a novel approach to compactly represent JSON-LD data using the 
Concise Binary Object Representation (CBOR) and bitmaps. Our proposed method is named as 
CBL which stands for CBOR, Bitmap and List of Key-value pairs. CBL leverages the ideas from CBOR and HDT to achieve 
an efficient encoding of JSON-LD data.
%In this paper, we study CBOR-LD scheme for encoding JSON-LD to a lightweight binary format CBOR (Compact Binary Object Representation). We also study possible optimisations such as using a custom dictionary. We provide a C library and evaluate it on several examples. 
Results demonstrate that our approach provides savings up to 95.1\% in terms of network overhead. This could be especially beneficial for IoT devices exchanging data over wireless networks. Moreover, our approach is more efficient than the current approach known as CBOR-LD, which is used to compact JSON-LD data. 

\end{abstract}

\section{Introduction}
Building upon the Internet of Things (IoT) paradigm, Web of Things (WoT) \cite{raggett2015web} integrates web 
technologies and knowledge graphs to facilitate the seamless interaction between devices. As 
a JSON-based format for representing and exchanging structured data online, JSON-LD~\cite{json-ld} can 
be leveraged by WoT for structured data exchange.

JSON-LD offers several advantages in the IoT and WoT domains, including enhanced semantic interoperability, streamlined data integration from disparate sources, contextualized 
information, and linked data. By utilizing semantic annotations, JSON-LD enables IoT devices and systems from various manufacturers to interpret data consistently, promoting seamless communication. Furthermore, JSON-LD's flexible and standardized format facilitates the 
integration of diverse data sources, such as sensors, actuators, and control systems. The inclusion of contextual information alongside data in JSON-LD enriches IoT applications by adding additional meaning and significance to the generated data.

Moreover, JSON-LD's linked data property enables the discovery and integration of distributed data sources, empowering more complex applications and analyses. This allows for flexible 
declarative search as data is represented in a graph format. For instance, queries can be formulated to search for rooms with lights turned ON in a building, or identify rooms with 
electric consumption exceeding a specific threshold. The graph structure also enables the inference of new facts through the use of reasoners. 

However, one disadvantage of JSON-LD is it's verbose format that can be a significant overhead for constrained WoT objects. Some of the constrained objects have limited bandwidth, small amount of memory, limited computational capabilities and run on a battery.  

This paper proposes a novel approach to compactly represent JSON-LD data using the Concise Binary Object Representation (CBOR) and bitmaps. Our proposed method is named as CBL (CBOR, Bitmap, and List). CBOR is a compact binary format that supports various data types for efficient memory usage. It is designed to be lightweight, easy to parse, and optimized for low-power devices. CBOR supports nested structures, and schema-based validation and facilitates readability through stream processing making it suitable for representing complex data in a compact format. HDT is another compact format designed for RDF data. CBL leverages the ideas from CBOR and HDT and optimizes them to achieve efficient encoding of JSON-LD data.

The contributions of this paper are as summarised follows: 
\begin{itemize}
    \item We propose a new encoding scheme called CBL\footnote{https://gitlab.com/coswot/cbl} (CBOR, Bitmap and List of key-value indices).
    \item JSON-LD data is encoded using a re-indexation dictionary of terms which is generated from each JSON-LD example, a bitmap indicating the JSON-LD structure, and a list of key-value indices pointing to the terms in the dictionary. Re-indexation dictionary is encoded using CBOR. A new encoding of map/array start and end is proposed using bitmap. This is followed by indices of key-value pairs encoded using a minimum number of bits.
    \item The encoding is optimized by considering ideas from CBOR-LD\cite{cbor-ld-24} and HDT as well as a simple but efficient way of encoding strings with fixed and variable parts.
\end{itemize}

%This paper contributes by proposing an open source C library\footnote{\url{https://gitlab.com/coswot/cborld-c}} to first encode JSON-LD terms to integers based on a dictionary and then to convert it to CBOR format.  
%C language was chosen to make it more suitable for embedded devices. Our implementation is based on an in-progress CBOR-LD standardisation in World Wide Web Consortium (W3C)\footnote{\url{https://json-ld.github.io/cbor-ld-spec/}}. 
Thus, in this paper, we also explore optimization possibilities such as encoding strings that have fixed and variable parts, etc. Our results show that CBL performs better than CBOR-LD, a recent method to compact JSON-LD data, and can obtain further savings in terms of bandwidth. 

This paper is organized as follows. Section II discusses background and related work. Section III details the proposed encoding scheme called CBL. Section IV performs evaluation on some example data and Section V concludes the paper.

\section{Background and Related work}
In this Section, we describe some background on JSON-LD and some related work for compacting  linked data. 
\subsection{JSON-LD}
\begin{figure}[!ht]
\vspace{0.05in}
    \centering
    \includegraphics[width = 0.45\textwidth]{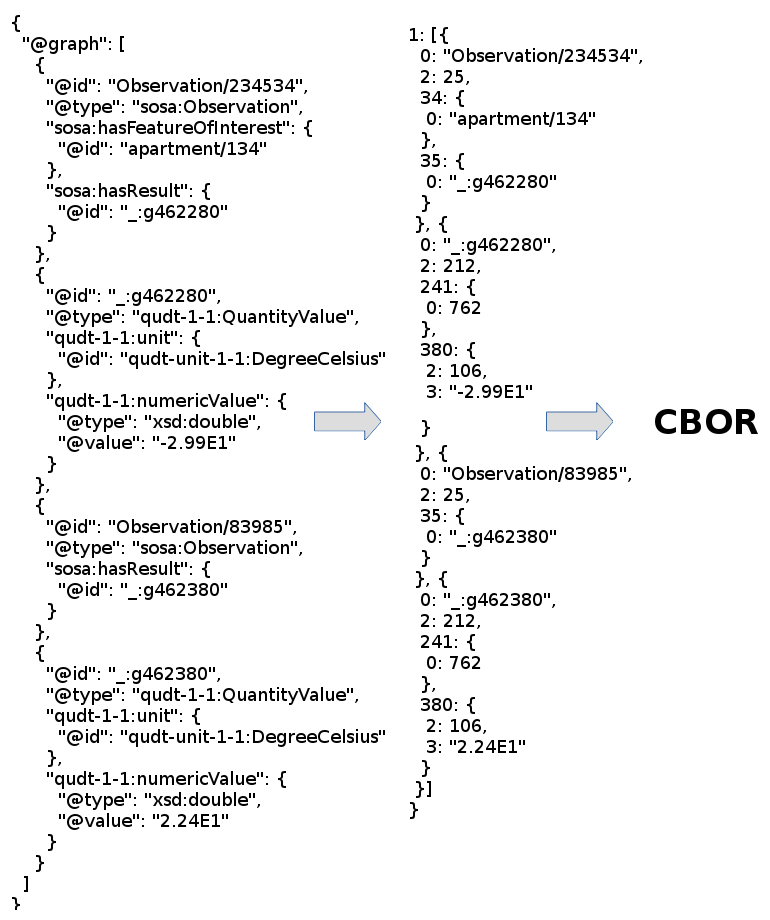}
    \vspace{-1em}
    \caption{JSON-LD to CBOR-LD \cite{raj2024poster}}
    \label{fig:cborld}
\vspace{-1em}
\end{figure}

Converting JSON-LD data to CBOR involves a series of pre-processing and transformation steps.
Figure~\ref{fig:cborld} illustrates an example from the SSN Ontology, known as Example 1 in 
the SSN document \cite{ssnw3c}. This example showcases a graph with nodes representing two observations
that demonstrate the difference between outside and inside temperatures, along with the 
absolute inside temperature.

The first observation is represented by the node with ID \texttt{Observation/234534}, which 
features an apartment identified as \texttt{apartment/134}. The observation also includes a result, 
denoted by the blank node \texttt{\_:g462280}, which represents a quantity value. This value is a 
temperature difference in Degree Celsius with a numerical value of ``-2.99E1'' (i.e., -29.9
°C).

%Converting JSON-LD data to CBOR involves pre-processing and transformation steps. 
%Figure~\ref{fig:cborld} illustrates an example from SSN Ontology, termed as Example 1 in the SSN document\cite{ssnw3c}. This example comprises a graph of nodes having 2 observations showing the difference between the outside and inside temperature, as well as the absolute inside temperature. First observation is represented by the node with the ID 
%\texttt{Observation/234534}. It has a feature of interest, which is an apartment 
%identified as \texttt{apartment/134}. The observation also has a result, which is a quantity value 
%represented by the blank node \texttt{\_:g462280}. The value of this quantity is a temperature difference with a unit of Degree Celsius and a numerical value of ``-2.99E1'' (i.e. -29.9 °C).
The second observation is represented by the node with ID \texttt{Observation/83985}. This observation also features a result, denoted by the blank node \texttt{\_:g462380}, which 
represents another quantity value. The value of this quantity is a temperature inside the apartment equal to ``2.24E1'' (22.4 °C). The distinction between whether the observation 
relates to the difference in temperatures or the temperature itself will be described by other ontology features.

%Second observation is represented by the node with the ID \texttt{Observation/83985}. This observation also has a result, which is another quantity value represented by the blank node 
%\texttt{\_:g462380}. The value of this quantity is a temperature inside the apartment equal to ``2.24E1'' ($22.4$ degrees Celsius). Whether the observation is about the difference in temperatures or the temperature itself will be described by other ontology features. 

Representing data in this manner offers several advantages, including improved 
interoperability, as previously discussed. It is evident that the knowledge about data and 
its semantics are embedded within the data itself, conveying what the data represents, as 
well as the relationships between data entities and standardized concepts.

However, this format can be verbose, resulting in significant bandwidth overhead for 
constrained objects. Therefore, we propose CBL as a solution to compact this data.

%Representing data like above has several advantages such as interoperability among others, as discussed before. It can be seen that the knowledge about data and semantics are represented in the data itself, i.e, what is data about, what is the relation of a data entity with other standardised concepts.

%However, this format can be verbose leading to significant bandwidth overhead for constrained objects. Thus, we propose CBL to compact this data.

\subsection{Related works on compact representations}
HDT (Header, Dictionary, and Triples)~\cite{fernandez2010compact} represents the state of the art on compacting the RDF (Resource Description Framework) data. RDF~\cite{mcbride2004resource} is a popular framework for representing linked data and provides formats such as Turtle. The authors in ~\cite{bazoobandi2015compact}
introduced a novel in-memory RDF dictionary using optimized Trie structures to efficiently compress common prefixes. 
The work in ~\cite{karim2021compact} proposes a compact
representation of semantic sensor data, in RDF format, by sending the repeated measurement values only once.

For JSON-LD, some previous works~\cite{charpenay2018towards} have also studied compact representations for JSON-LD as well as RDF data.  The work in ~\cite{charpenay2018towards} compared different approaches and proposed JSON-LD compaction by mapping resource identifiers to shorter strings. They found that such compaction coupled with CBOR can lead to compaction ratios around 50-60\% compared to HDT. 

More recently, the work on using CBOR to compact the JSON-LD data has been ongoing in W3C named as CBOR-LD. Thus, we compare our approach CBL with CBOR-LD in this paper and show that CBL is more efficient. We find that the HDT format can be further optimized for constrained WoT objects and as it works mainly for RDF, it needs to be adapted for JSON-LD. In this paper, we optimize and adapt HDT by combining ideas from CBOR-LD. The resulting idea called CBL is shown to be better than CBOR-LD. 

% \lstinputlisting{ssn-example-1.jsonld}
% \lstinputlisting{intermediate-conversion}

\subsection{CBOR-LD}
This section explains the background on CBOR-LD. In CBOR-LD \cite{cbor-ld-24}, first the standard JSON-LD terms like \texttt{@id, @graph, @type} are mapped to integers like $0, 1, 2$, etc. assuming a universal dictionary which can be standardized. Note that this mapping is subject to change as the standardization is in process. This mapping to integers is important as CBOR is more efficient for encoding integers as compared to text. Next, the standard terms from different ontologies like SSN, SOSA, and QUDT, (i.e. the ones used in the example) are mapped to integers assuming that these terms in the ontology can be obtained from the given ontology, sorted, and then mapped one by one to integers. As ontologies are standardized and their terms are well known, if this above algorithm of constructing the dictionary is standardized as well then any application can construct this dictionary independently. In JSON-LD,  the key \texttt{@context} is used to define the mapping between different terms and their corresponding IRIs (Internationalized Resource Identifiers). This mapping can be directly present in the data or can be referenced using a URL. Such mapping can be used to construct the dictionary. Thus, all terms from different ontologies like \texttt{sosa:observation}, which is a string of size $16$ bytes, are transformed to integer values such as $25$ which takes only $1-2$ bytes. Timestamps and dates like \texttt{2017-04-16T00:00:12+00:00} are also converted to Unix time epochs as integers. Finally, the graph consisting of integers is then serialized into CBOR format which finally compacts the size of the graph.

%Once a graph is converted to CBOR then some processing operations, querying partial graph extraction, etc., can always be applied to it without converting it back to the original format. Thus, such data can stay in CBOR format inside the network of constrained objects and save bandwidth during exchanges. That way the need to decode it will arise only during some final operations in the application. 

%\subsection{Custom dictionary}
Note that after the CBOR-LD encoding, as described above, some terms will still be left that could not be encoded as integers. This is because there may be data-specific, non-standard, terms such as \texttt{apartment/134/electricConsumption}, \texttt{sensor/926} as well as some base URLs. Nevertheless, such terms may also stay fixed for a given application and can thus be encoded.  Work in \cite{raj2024poster} proposed to use a custom dictionary for such fixed terms and base URLs. Such a custom dictionary will be specific to the application and for example, can be exchanged in an offline manner. %For memory optimization, only a sub-dictionary containing the required terms by the IoT object may be used. Finally, the resulting CBOR data can in some cases be compressed a bit more by using compression methods and tools such as gzip. 
We will also compare CBL with this version which we call CBOR-LD-OPT.

\section{CBOR, Bitmap, and List (CBL)}

We present a novel approach to compactly represent JSON-LD data using the Concise Binary Object Representation (CBOR), bitmaps and list of key-value indices. Our proposed method, referred to as CBL, leverages the ideas from CBOR, CBOR-LD and HDT to achieve efficient encoding of JSON-LD data. 

We will focus on $6$ examples from SSN ontology document \cite{ssnw3c} numbered according to W3C recommendation: Example $1, 10, 12, 14, 17$ and $19$. They are about the Indoor and Outdoor Temperature, electric consumption of an apartment, sensor used to observe tree height, observation of seismograph, movements of spinning cups on wind sensor, and CO2 level observed in an ice core, as in Table~\ref{tab:SSN-examples}. We will also use the example SSN-1 as a running example for illustration.

In CBL, first of all, the idea of CBOR-LD to map standard terms to integers is used. A static dictionary is used which is not sent but negotiated offline because it mostly contains standard and fixed terms. The static dictionary also consists of a mapping of standard JSON-LD keywords to integers. This is done using the approach from Section II C: basic standard JSON-LD terms like \texttt{@id, @graph, @type}, which are mapped to integers like $0, 1, 2$, etc. in the static dictionary.  This mapping is shown in Table~\ref{tab:terms}. Additionally, other standard JSON-LD keywords are included if  they are used in the data. Next, the standard terms from different ontologies like SSN, SOSA, and QUDT, (i.e. the ones used in the example) are mapped to integers and so on.

%This dictionary includes basic standard JSON-LD terms like \texttt{@id, @graph, @type}, which are mapped to integers like $0, 1, 2$, etc. in the static dictionary.  This mapping is shown in Table~\ref{tab:terms}. Additionally, other standard JSON-LD keywords are included if  they are used in the data.

%Next, the standard terms from different ontologies like SSN, SOSA, and QUDT, (i.e. the ones used in the example) are mapped to integers assuming that these terms in the ontology can be obtained from the given ontology, sorted and then mapped one by one to integers. As ontologies are standardized and their terms are well known, if this above algorithm of constructing the dictionary is standardized as well then any application can construct this dictionary independently. In JSON-LD,  the key \texttt{@context} is used to define the mapping between different terms and their corresponding IRIs (Internationalized Resource Identifiers). This mapping can be directly present in the data or can be referenced using a URL. Such mapping can be used to construct the dictionary. Thus, all terms from different ontologies like \texttt{sosa:observation}, which is a string of size $16$ bytes, are transformed to integer values such as $25$ which takes only $1-2$ bytes. Timestamps and dates like \texttt{2017-04-16T00:00:12+00:00} are also converted to Unix time epochs as integers. 
%Finally the graph consisting of integers is then serialized into CBOR format which finally compacts the size of the graph.

\begin{table}
{
\centering
   % \begin{center}
     \begin{tabular}{|l|c|}
        \hline
        Term & Encoded as Integer\\
         \hline
         @id &0\\
        @graph &1\\
        @type &2\\
        @value &3\\
        @context &4\\
        @language& 5\\
    \hline
    \end{tabular}
    \vspace{1em}
    \caption{Standard and most frequent JSON-LD terms encoded as integers.}
    \label{tab:terms}
 %  \end{center}
 }
\vspace{-2em}
\end{table}

\begin{figure*}[!ht]
%\vspace{0.05in}
    \centering
    \includegraphics[width = 0.7\textwidth]{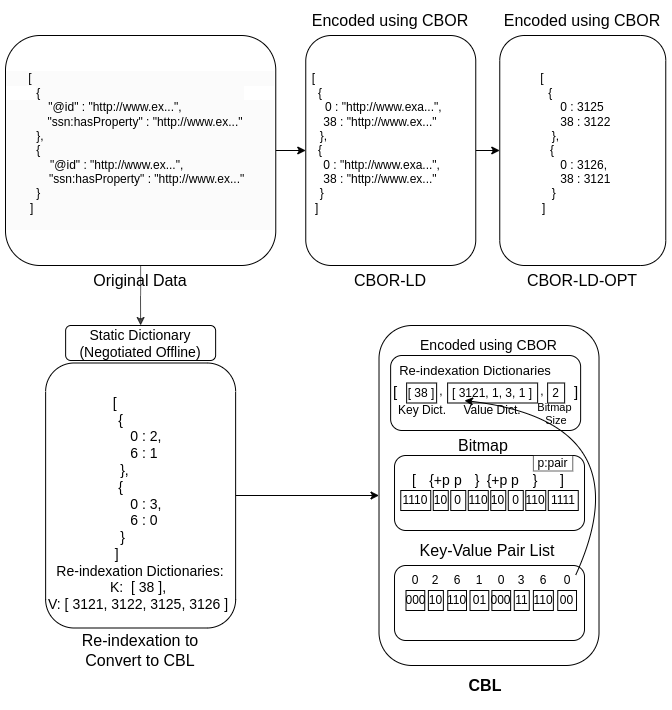}
    \vspace{-2em}
    \caption{CBL approach as compared to CBOR-LD approach. For CBL, note that key-value indices below point to items in the dictionary, such as shown for value index 0 pointing to 3121. Also the key indices from 0 to 5 are implied to be @id, @value.. etc. and hence are not even sent in the  dictionary.}
    \label{fig:cbl}
\vspace{-1em}
\end{figure*}

%Other terms from ontologies such as SSN and SOSA are also mapped to integers like in CBOR-LD. 
Next, only the Re-indexation dictionary, bitmap, and a list of key-value pairs are sent in the CBL encoded data. The three parts of the encoded data are described below and the general method of CBL vs. CBOR-LD is described in Figure~\ref{fig:cbl}.

\subsection{CBOR Encoded Re-indexation Dictionary}
\emph{The purpose of the re-indexation dictionary is to re-map the terms being used in the transmitted data to smaller integers.} Encoding smaller integers/indices takes less number of bits ($log_2(maxIndex)$). Whereas, the static dictionary indexes all the possible terms of ontologies being used, and their indices can become high, needing many bytes.

Next, this re-indexation dictionary is encoded using CBOR. It contains terms such as integers, strings, arrays, and 
other data types. By reusing an existing standard CBOR, we benefit from its well-established 
encoding schemes and avoid introducing new complexities.

First, there will be an array of keys and then values. Their index in the array will be implied and thus will not need to be transmitted. 
Now, thanks to this arrangement, the integers and remaining strings can be sorted and we find that {\bf delta encoding} can be efficiently applied. For example in case the key array is [34, 35, 37 ..] then we can just transmit [34, 1, 2 ..]. This has the advantage that the smaller numbers will take a lesser number of bytes with CBOR. Thus, the keys and values array in the re-indexation dictionary are sent after applying delta encoding.
%\emph{The idea of the dynamic dictionary is to re-index terms that are present in the current graphs.} That way there index will be a smaller number and hence CBL will be able to encode it using less number of bits.

{\bf Keys:} Taking an example of SSN-1: if @id, @type, sosa:hasFeatureOfInterest, sosa:hasResult, etc. are the keys present in the current graph being transmitted. They will be encoded as 0, 2, 34, 35, 241, and 380 respectively by the static dictionary. So the re-indexation dictionary will contain an array of keys with sorted terms [34, 35, 241, 380] because @id and @type are encoded always as 0 and 2 by CBL and are thus implied. The first five indices 0 to 5 are reserved for such frequent terms shown in Table~\ref{tab:terms}. The remaining keys  sosa:hasFeatureOfInterest, sosa:hasResult 
were encoded as 34 and 35 in the static dictionary and are now re-indexed starting from index 6. Finally, delta encoding is applied and the keys array, starting from $6^{th}$ element, becomes [34, 1, 206, 139].

{\bf Values:} Next there will be an array containing value terms. As CBOR encoding includes the type values for standard items such as arrays, the decoder will be able to know about the start of the next array. Values terms can be standard terms such as sosa:Observation, strings with fixed part and a variable part such as ``Observation/234534'' or values such as ``-29.9'', etc. We propose different strategies to encode them. Standard terms like sosa:Observation are encoded as integers according to the static dictionary and note that these terms in the current graph will be re-indexed. 

To be able to apply delta encoding and to be able to differentiate which is which, the elements appear in the following order: original arrays of data values if present in the data, then integers and finally strings. This is to avoid confusion because we will see strings are also encoded using arrays to perform delta coding.

%{\bf Delta encoding:} 
For strings with fixed and variable parts such as  ``Observation/234534'', the part ``Observation/'' will stay fixed but the observation number may change. We can first include the string and then use delta encoding to next occurrences of similar strings. Otherwise, we can also encode the first occurrence as an integer in the static dictionary, let's say 900, and then the variable part is encoded as an array. In the array the first integer if -ve means that we take the previous string, remove the indicated number of characters, and then append a number indicated by the 2nd element. For example there are `\_:g462280' and `\_:g462380'. Using delta encoding `\_:g462380' can be encoded as [-3, 380], which says that we remove the last 3 characters from the previous string and then append 380 at the end. Such encoding can also be beneficial for base URLs where only the last part is variable. Numerical values such as ``-29.9'' are either encoded as corresponding floats or left as strings in this case as float will be bigger than the string. In case a string indexed by an integer needs to be appended then a CBOR tag like tag 25 can be used which references to previous strings.

For the SSN-1 example, the values array will be initialized as follows:
[25, 106, 212, 762, 3100, `Observation/234534', `Observation/83985',`-29.9', `22.4', `\_:g462280', `\_:g462380']. Then after delta encoding, it will finally be sent as:
[25, 81, 106, 550, 2338, `Observation/234534', [-6, 83985], `-29.9', `22.4', `\_:g462280', [-3, 380]]. The above saves 16 bytes after CBOR conversion. For further savings we encode `Observation/234534' as integer in static dictionary and further observations are encoded using delta encoding. %Note that `\_:g' which is the initial part of the blank node identifier could have been in the static dictionary. However, for a fair comparison with CBOR-LD, we leave it in the re-indexation dictionary.

\subsection{Bitmap}

First a length indicator encoded using CBOR is present to indicate the size of the following bitmap. A bitmap is used to represent the curly braces '\{', '\}', square brackets '[', ']' (i.e. start and end of maps and arrays), presence of key-value pairs or single array elements. The bitmap will indicate where to fill the key-value pairs using the list of key-value indices which follows later. This approach allows for efficient encoding of these common data structures. We use an encoding based on Huffman coding such that the frequently occurring symbols are encoded using lesser number of bits. The pairs and maps are given less number of bits as they are more frequent as compared to the arrays. The bit-codes for the start and ending symbols are provided in Table~\ref{tab:bitcodes}. Note that 10 represents the start of the map plus the first key-value pair.

As an example, if the data was \{380:\{2:106, 3:``2.24E1''\}\} then the bitmap will encode the structure as 10 10 0 110 110. Later during decoding, the terms, whose indices are present in the list of key-value pairs, will be filled in the places indicated by the bit-code of key-value pairs. 

\begin{table}
{
\centering
   % \begin{center}
     \begin{tabular}{|c|c|}
        \hline
        Symbol & Bitcode  \\
         \hline
         key-value pair or an element of an array & 0\\
         start map '\{' + a first pair (assuming no empty maps) & 10\\
end map '\}' & 110\\
start array '[' & 1110\\
end array ']' & 1111\\
    \hline
    \end{tabular}
    \vspace{1em}
    \caption{Bitcodes for array/map/pair start and begin.}
    \label{tab:bitcodes}
 %  \end{center}
 }
\vspace {-2em}
\end{table}

\subsection{List of Key-Value Pair Indices}
Re-indexation dictionary was used to re-index the terms so that they can be represented using a fewer number of bits. Now, this list contains the new indices of the keys and values (including single array elements) following the same structure and the same order as they occur in the original data and thus the bitmap. These indices are like pointers to the terms present in the re-indexation dictionary and thanks to re-indexation these pointers can be represented using a very few bits. 

Key-value pairs are encoded using $k = \lceil log_2(K) \rceil$ bits to represent the key indices, where $K$ is the max index of the keys re-indexation dictionary and $v= \lceil log_2(V) \rceil$ bits to represent the 
value indices, where $V$ is the max index of values re-indexation dictionary (CBOR encoding of arrays has the array size information). Re-indexation of terms reduces the value of the maximum index and hence the number of bits required. For example, indices up to 7 can be represented using just 3 bits, or indices up to 15 need just 4 bits, and so on.
After writing the list of key-value indices, if 
the remaining bits are needed to complete a whole byte, they are padded with zeros. 

Taking the example of SSN-1, the list of key-value indices will be: 1,0,5,2,0 ..., encoded using $k$ and $v$ bits for keys and values. Key and value indices in the case of SSN-1 will both need just 4 bits. The data structure of SSN-1 is indicated by the Bitmap:  implying that the value associated to the key ``1'' or ``@graph'' is actually an array of several maps.  Then 0:5, 2:0 are key-value pairs, etc. In the above key 0 means ``@id'', key 2 means ``@type'',  the string ``Observation/234534'' included in values array (please see the end of Section III A) of re-indexation dictionary got re-indexed as 5, sosa:Observation encoded as 25 got re-indexed as 0 in the value dictionary. 

\section{Evaluation}

In this section, we provide a detailed analysis of our proposed method CBL, 
including its performance evaluation and comparison with existing CBOR-LD approach.

\begin{table}
{
%\vspace{0.13in}
\centering
   % \begin{center}
     \begin{tabular}{|c|c|}
        \hline
         JSON-LD Data & Description \\
         \hline
         SSN Example 1 & Indoor and Outdoor Temperature \\
          SSN Example 10 & Electric consumption of an apartment \\
          SSN Example 12 & Sensor used to observe tree height \\
          SSN Example 14 &Observation of seismograph \\
          SSN Example 17 & Movements of spinning cups on wind sensor \\
          SSN Example 19 & CO2 level observed in an ice core  \\     
         \hline
    \end{tabular}
    \vspace{1em}
    \caption{SSN Examples used for evaluation.}
    
    \label{tab:SSN-examples}
 %  \end{center}
 }
 \vspace{-3em}
\end{table}

The 6 examples were taken from SSN document~\cite{ssnw3c} for this study and they were converted to JSON-LD using RDF distiller\footnote{http://rdf.greggkellogg.net/distiller?command=serialize}. The URLs of different ontologies which are provided by the key \texttt{@context} were removed as they are well-known URLs.
\begin{table*}
{
\vspace{0.13in}
\centering
   % \begin{center}
     \begin{tabular}{c|c|c|c|c|c|c}
         JSON-LD Data &  Size & Gzip& CBOR-LD & CBOR-LD-OPT & CBOR-LD-OPT+Gzip &\bf{CBL}\\
         & Bytes &Bytes& Bytes, Savings & Bytes, Savings &Bytes, Savings&Bytes, Savings\\
         \hline
         SSN Example 1 &  904 &301& 178, 80.3\% & 136,  84.9\%& 118, 86.9\%& \bf{84, 90.7\%}\\
          SSN Example 10 &  5322 &830& 1616, 69.6\% & 510, 90.4\%& 340, 93.6\%&\bf{323, 93.9\%}\\
          SSN Example 12 &  3748 &573& 1069, 71.5\% & 519, 86.2\%& 314, 91.6\%&\bf{312, 91.7\%}\\
          SSN Example 14 & 3503 &773& 1156, 67\% & 480, 86.3\%& 390, 88.9\%&\bf{363, 89.6\%}\\
          SSN Example 17 &  2817 &476& 1057, 62.5\% & 227, 91.9\%& 166, 94.1\%&\bf{138, 95.1\%}\\
          SSN Example 19 &  2414 &535& 754, 68.8\% & 209, 91.3\%& 194, 91.9\%&\bf{149, 93.8\%}\\     
         \hline
    \end{tabular}
    \vspace{1em}
    \caption{Comparison of CBL with CBOR-LD. Note that \texttt{@context} was omitted. Also the comments present in the graphs were omitted as sending big comments over constrained networks will be inefficient.}
    \label{tab:performance}
    \vspace{-2em}
 %  \end{center}
 }
%\vspace{-2em}
\end{table*}
\begin{table*}
{
\vspace{0.13in}
\centering
   % \begin{center}
     \begin{tabular}{c|c|c|c|c}
         JSON-LD Data &  CBL & CBL+Gzip& CBL-CBOR & CBL w/o Delta encoding\\
         & Bytes &Bytes& Bytes & Bytes\\
         \hline
         SSN Example 1 & 84 &114&106&94\\
          SSN Example 10 &323 &352&418&393\\
          SSN Example 12 &312&335&398&413\\
          SSN Example 14 &363&395&430&418\\
          SSN Example 17 &138&165&183&164  \\
          SSN Example 19 &149&183&186&179  \\     
         \hline
    \end{tabular}
    \vspace{1em}
    \caption{Ablation studies with different CBL variants}
    \label{tab:ablation}
 %  \end{center}
 }
   \vspace{-2em}
%\vspace{-2em}
\end{table*}
Table~\ref{tab:performance} shows the results. We can see that CBL can reduce the size of data significantly. We estimate savings as $100\cdot\frac{\texttt{Original size - Encoded size}}{\texttt{Original size}}$. 

We can see that CBOR-LD can provide savings ranging from $62\%$ to $80\%$ approx. Some graphs like Example $14$ could not reach higher savings because they have several specific URLs. 
The work in \cite{raj2024poster} used a custom dictionary by encoding some more terms like base URLs as integers. We call this version CBOR-LD-OPT which provides better savings as compared to the default CBOR-LD algorithm. The savings are between $85\%$ to $92\%$ savings with the examples considered.
%There is still some more scope for improvement. For example, the blank node identities like \texttt{\_:g462380} were not encoded and the ideas discussed before can be explored. 
%Moreover, we encoded the specific URLs as integers, but a more general or hierarchical way of encoding URLs may be found for example using arrays. 
Gzip can also help in compacting the data as well and CBOR-LD-OPT+Gzip obtains savings up to $94.1\%$.  However, gzip sometimes takes more space than the original small data because of overheads. Hence, the resulting size should be checked before sending the data over the network. 

Finally, CBL shows the best performance on all examples and it performs better than even CBOR-LD-OPT combined with gzip. With CBL there is no need to use gzip which may not always be available on constrained nodes. With CBL constrained IoT nodes can significantly reduce their network overhead and savings are between 89.9\% to 95.5\% for the examples considered.
We can also compare relative performances by taking the CBOR-LD-OPT or CBOR-LD-OPT+Gzip size as the original size in the savings formula. We find that CBL relatively saves between 24.4\% to 39.9\% with respect to CBOR-LD-OPT and between 0.6\% to 28.8\% with respect to CBOR-LD-OPT+Gzip. We also tested original HDT, but it took higher number of bytes (not reported in the Table) as it is not adapted for small IoT data.

\subsection{Ablation Studies}
In this Section, we compare different possible variants of CBL encoding in Table~\ref{tab:ablation}. CBL+Gzip corresponds to the case when gzip is applied after CBL. In this case, the resulting size is even larger than before because CBL is already very compacted and Gzip does not work well with small-sized data. Thus, we recommend to use CBL without Gzip. CBL-CBOR is the variant when after the re-indexation dictionary, instead of bitmap and list, the CBOR encoded map is sent with integers replaced with re-indexed ones. This variant takes more space as CBOR has some extra overhead in terms of types and it needs at least 2 bytes to represent any integer more than 23. 

CBL wo Delta encoding variant shows CBL without delta encoding and we can see without delta encoding some savings in terms of bytes are lost. %Finally, we compare the variant CBL-HDT which encodes the bitmap and triples according to HDT. This variant is efficient, but after decoding will produce RDF triples and not JSON-LD. A step to convert it back to JSON-LD will be needed.

\section{Discussion}
We found that CBL outperformed the CBOR-LD approach. This is mainly because CBOR-LD does not remove redundancy in the form of repeating terms. With CBL, if some terms repeat then they simply appear once in the dictionary and each occurrence is pointed by the smaller size of index present in key-value indices list. 
Re-indexation dictionary can be sorted, which allows delta encoding to work efficiently.

We proposed basic delta encoding strategies for integers and strings and perhaps better strategies are possible. For example some times whole maps may repeat and currently we do not eliminate such redundancies. After looking at the data, we found that to be the case for SSN-12 example. This could be the reason that CBOR-LD + Gzip size for SSN-12 is similar to CBL. This is because Gzip can eliminate different types of redundancies even though it is not so efficient for small data. 

CBL can be useful for WoT or IoT networks which want to exploit linked data capabilities. CBL can be exchanged inside the network and will support declarative search, reasoning, contextualised information etc., over compact data and hypothetically will also provide savings in terms of memory.

\section{Conclusion}
This paper proposed CBL method which offers several advantages over existing approaches. By using CBOR as the base encoding scheme, we inherit its efficiency and flexibility. The bitmap representation of pairs provides a compact way to encode JSON-LD data structures. Finally, the use of a static dictionary allows for offline sharing. The proposed approach offers significant savings in terms of network overhead as compared to current approaches like CBOR-LD, making it an ideal solution for constrained WoT objects. Future 
work includes further optimizations for repetitive text or URL encoding, decreasing further the size of the dynamic dictionary transmitted, and testing in real-world scenarios.

\section{Acknowledgements}
This work is supported by grant ANR-19-CE23-0012 from the Agence Nationale de la Recherche, France, for the CoSWoT project\footnote{\url{https://coswot.gitlab.io/}}. We also thank CoSWoT project colleagues like Victor Charpenay for their constructive comments.
%This work was supported by Agence Nationale de la Recherche, France for project CoSWoT\footnote{\url{https://coswot.gitlab.io/}}, grant ANR-19-CE23-0012.

\bibliographystyle{IEEEtran}
\bibliography{biblio}
\end{document}